\documentclass{PHYEAUTH}
\usepackage[dvips]{graphicx}
\usepackage{amsmath}
\usepackage{amssymb}

\begin{document}

\begin{frontmatter}

\title{Theory of Striped Hall Ferromagnets}

\author[address1]{Nobuki Maeda\thanksref{thank1}},

\address[address1]{Department of Physics, Hokkaido University, 
Sapporo 060-0810, Japan}

\thanks[thank1]{
Corresponding author. 
E-mail: maeda@particle.sci.hokudai.ac.jp}

\begin{abstract}
We study spin and charge striped states at the half-filled high Landau 
level in the zero Zeeman energy limit using a Hartree-Fock approximation. 
It is shown that a ferromagnetic striped Hall state is more stable 
than the antiferromagnetic striped state or charge striped state. 
We calculate the collective excitations using the single mode 
approximation. 
\end{abstract}

\end{frontmatter}


Recently, striped states have been proposed for the anisotropic 
states in the quantum Hall system at several filling factors. 
At half-filled high Landau levels, highly anisotropic states 
are observed\cite{Lilly,Du} and explained by the anisotropic charge 
density wave or charge striped state.\cite{Kou,Moe} 
Furthermore at even integer fillings, highly anisotropic states 
have been observed in a quantum well system.\cite{Pan} 
A kind of the striped state or domain wall 
related to spin or pseudospin is a candidate of the 
anisotropic state.\cite{Dem,Rez} 
Considering the spin and pseudospin degree of freedom, a very rich 
structure has been predicted theoretically.\cite{Jun,Wan,Cote} 

In this paper we consider a possibility of spin and charge striped 
states in an ideal 2D electron system at half-filled high Landau 
level in the zero Zeeman energy limit. 
Using the Hartree-Fock approximation, it is shown that 
a ferromagnetic striped Hall state has a lower energy 
than the antiferromagnetic striped state or charge striped state. 

We calculate the collective excitations using the single mode 
approximation. 
There are two kinds of the excitations, namely the phonon 
and spin wave due to the spontaneous breakdown of the translational 
symmetry and spin rotational symmetry, respectively. 
It is shown that 
the dispersion of the spin wave has a weaker anisotropy than 
the phonon dispersion. 

Let us consider the 2D electron system in a perpendicular 
strong magnetic field. 
The kinetic energy is quenched in the Landau level space and 
neglected in this paper. 
Then we consider only the following interaction Hamiltonian
\begin{equation}
H_{\rm int}=:{1\over2}\int{d^2k\over(2\pi)^2}
\rho({\bf k})V(k)\rho(-{\bf k}):,
\end{equation}
where $V(k)=2\pi q^2/k$ and $\rho$ is the density operator. 
The Zeeman energy term is neglected. 
We use the unit $\hbar=c=1$ and set $a=\sqrt{2\pi/eB}=1$. 
In the von Neumann lattice formalis\cite{Von}, 
the density operator $\int d^2x e^{i{\bf k}\cdot
{\bf x}}\sum_\alpha\psi_\alpha^\dagger({\bf x})\psi_\alpha({\bf x})$
is written in the projected $l$ th Landau level as
\begin{eqnarray}
\rho({\bf k})&=&
f_l(k)\sum_\alpha\int_{\rm BZ}{d^2p\over(2\pi)^2}b_\alpha^\dagger({\bf p})
b_\alpha({\bf p}-\hat{\bf k})\nonumber\\
&&\times e^{-{i\over4\pi}{\hat k}_x(2p_y-{\hat k}_y)},
\end{eqnarray}
where $f_l(k)=e^{-k^2/8\pi}L_l(k^2/4\pi)$, $\hat k=(r k_x,k_y/r)$, 
$\alpha=\uparrow$, $\downarrow$, and $r$ is an asymmetric parameter of 
the unit cell of von Neumann lattices. 
In the mean field theory of the striped state, the parameter $r$ 
becomes the period of the stripe. 
The spin density operators projected to the $l$ th Landau level 
are written by
\begin{eqnarray}
s_i({\bf k})&=&
f_l(k)\sum_\alpha\int_{\rm BZ}{d^2p\over(2\pi)^2}b_\alpha^\dagger({\bf p})
{\sigma^i_{\alpha\beta}\over2}
b_\beta({\bf p}-\hat{\bf k})\nonumber\\
&&\times e^{-{i\over4\pi}{\hat k}_x(2p_y-{\hat k}_y)},
\end{eqnarray}
where $\sigma_i$ is the Pauli matrix and $i=$1, 2, 3. 
Ladder operators are defined by $s_{\pm}=s_1\pm is_2$. 

In the Hartree-Fock approximation, we consider the following mean 
field
\begin{eqnarray}
\langle b^\dagger_\beta({\bf p})b_\alpha({\bf p}')\rangle
&=&U_{\alpha\beta}({\bf p})\\
&&\times\sum_N(2\pi)^2\delta({\bf p}-{\bf p}'+2 \pi{\bf N})
e^{i\phi(p,N)},\nonumber
\end{eqnarray}
where $U_{\alpha\beta}({\bf p})$ is periodic function in the 
magnetic Brillouin zone and $\phi(p,N)=\pi(N_x+N_y)-N_y p_x$. 
Using the Pauli matrix, $U_{\alpha\beta}({\bf p})$ is written as 
\begin{equation}
U_{\alpha\beta}({\bf p})=U_0({\bf p}){\delta_{\alpha\beta}\over2}+
\sum_i U_i({\bf p}){\sigma^i_{\alpha\beta}\over2}.
\end{equation}
$U_0$ and $U_i$ are distributions of the number density and 
spin density in the momentum space, respectively. 
Using the mean field, Hartree-Fock Hamiltonian reads
\begin{eqnarray}
H_{\rm HF}&=&\sum_{\alpha\beta}\int_{\rm BZ}{d^2 p\over(2\pi)^2}
b_\alpha^\dagger({\bf p})\epsilon_{\alpha\beta}({\bf p})
b_\beta({\bf p})\label{HF}\\
&&-{1\over2}\sum_{\alpha\beta}\int_{\rm BZ}{d^2 p\over(2\pi)^2}
\epsilon_{\alpha\beta}({\bf p})U_{\alpha\beta}({\bf p})
(2\pi)^2\delta({\bf p}=0).
\nonumber
\end{eqnarray}
The one-particle energy $\epsilon_{\alpha\beta}({\bf p})$ is given by
\begin{eqnarray}
\epsilon_{\alpha\beta}({\bf p})&=&\int_{\rm BZ}{d^2p'\over(2\pi)^2}
[\delta_{\alpha\beta}v_{\rm H}({\bf p}-{\bf p}')
{\rm tr}(U({\bf p}'))\\
&&\qquad -v_{\rm F}({\bf p}-{\bf p}')U_{\alpha\beta}({\bf p}')].
\nonumber
\end{eqnarray}
The Hartree potential $v_{\rm H}$ and Fock potential $v_{\rm F}$ 
are given by
$
v_{\rm H}({\bf p})=\sum_N v_l(2\pi\tilde{\bf N})e^{-iN_y p_x+iN_xp_y},\ 
v_{\rm F}({\bf p})=\sum_N v_l(\tilde{\bf p}+2\pi\tilde{\bf N}),
$
where $v_l({\bf p})=f_l(p)V(p)$ and ${\tilde p}=(p_x/r,r p_y)$. 
In the followings we present three self-consistent Hartree-Fock 
state at $\nu=2 l+1/2$. 

\begin{figure}[t]
\begin{center}\leavevmode
\includegraphics[width=0.6\linewidth]{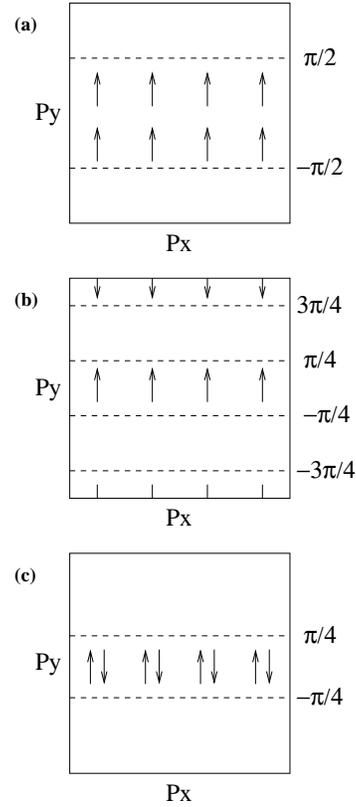}
\caption{ Schematic spin configurations in the Brillouin zone for the 
ferromagnetic striped state (a), antiferromagnetic striped state (b), 
and charge striped state (c). 
Dotted lines stand for Fermi surfaces. 
}\label{figurename}\end{center}\end{figure}

(a) {\it Ferromegnetic striped state}: 
First we consider the ferromagnetic striped state, in which 
the spin up state is filled periodically in 
$x$ direction and spin down state is completely empty. 
The period of the stripe is given by parameter $r$. 
The corresponding Fermi sea is a stripe extending in $p_x$ 
direction.\cite{Imo,SMA} 
See Fig.~1 (a). 
The corresponding mean field is
$
U_0({\bf p})=\theta(\pi/2-\vert p_y\vert),\ 
U_i({\bf p})=(0,0,\theta(\pi/2-\vert p_y\vert))$. 
Using this mean field, the one-particle energy is obtained as
\begin{equation}
\epsilon_{\alpha\beta}({\bf p})=
\left(\begin{array}{cc}
\epsilon_\uparrow(p_y)&0\\
0&\epsilon_\downarrow(p_y)
\end{array}\right),
\end{equation}
where 
\begin{eqnarray}
\epsilon_\uparrow(p_y)&=&\int_{-\pi/2}^{\pi/2}{dp'_y\over2\pi}
\int_{-\pi}^\pi{dp'_x\over2\pi}v_{\rm HF}({\bf p}-{\bf p}'), \nonumber\\
\epsilon_\downarrow(p_y)&=&0,
\end{eqnarray}
where $v_{\rm HF}=v_{\rm H}-v_{\rm F}$. 
The energy per particle $E_f(r)$ is given by
\begin{equation}
E_f(r)=\int_{-\pi/2}^{\pi/2}{dp_y\over2\pi}\epsilon_\uparrow(p_y)
\end{equation}

(b) {\it Antiferromagnetic striped state}: 
Next we consider an antiferromagnetic striped state in which 
the spin and charge density is periodic in $x$ direction and 
uniform in $y$ direction. 
The corresponding Fermi sea is two stripes extending in $p_x$ 
direction. 
See Fig.~1 (b). 
Then the mean field for the antiferromagnetic striped state is given by
$
U_0({\bf p})=\theta(\pi/4-\vert p_y\vert)+
\theta(\vert p_y\vert-3\pi/4),\ 
U_i({\bf p})=\theta(\pi/4-\vert p_y\vert)-
\theta(\vert p_y\vert-3\pi/4)$. 
Using this mean field, the one-particle energy is obtained as
\begin{equation}
\epsilon_{\alpha\beta}({\bf p})=
\left(\begin{array}{cc}
\epsilon_\uparrow(p_y)&0\\
0&\epsilon_\downarrow(p_y)
\end{array}\right),
\end{equation}
where 
\begin{eqnarray}
\epsilon_\uparrow(p_y)&=&\left(\int_{-\pi}^{-3\pi/4}+\int_{3\pi/4}^{\pi}
\right){dp'_y\over2\pi}\int_{-\pi}^\pi{dp'_x\over2\pi}
v_{\rm H}({\bf p}-{\bf p}')
\nonumber\\
&&+\int_{-\pi/4}^{\pi/4}{dp'_y\over2\pi}\int_{-\pi}^\pi{dp'_x\over2\pi}
v_{\rm HF}({\bf p}-{\bf p}'), 
\nonumber\\
\epsilon_\downarrow(p_y)&=&\epsilon_\uparrow(p_y-\pi).
\end{eqnarray}
The energy per particle $E_a(r)$ is given by
\begin{eqnarray}
E_a(r)&=&\int_{-\pi/4}^{\pi/4}{dp_y\over2\pi}\epsilon_\uparrow(p_y)\\
&+&\left(\int_{-\pi}^{-3\pi/4}+\int_{3\pi/4}^{\pi}\right)
{dp_y\over2\pi}\epsilon_\downarrow(p_y).
\nonumber
\end{eqnarray}

(c) {\it Charge striped state}: 
Finally we consider the charge striped state in which 
the spin up states and down states are filled 
periodically in $x$ direction and uniformly in $y$ direction. 
In this state, translation symmetry in $x$ direction is broken. 
The corresponding Fermi sea is a stripe extending in $p_x$ 
direction. 
See Fig.~1 (c). 
Then the mean field for the spin striped state is given by
$
U_0({\bf p})=2\theta(\pi/4-\vert p_y\vert),\ U_i({\bf p})=0$. 
Using this mean field, the one-particle energy is obtained as
\begin{equation}
\epsilon_{\alpha\beta}({\bf p})=
\left(\begin{array}{cc}
\epsilon_\uparrow(p_y)&0\\
0&\epsilon_\downarrow(p_y)
\end{array}\right),
\end{equation}
where 
\begin{eqnarray}
\epsilon_\uparrow(p_y)&=&\int_{-\pi/4}^{\pi/4}{dp'_y\over2\pi}
\int_{-\pi}^\pi{dp'_x\over2\pi}\{2v_{\rm H}({\bf p}-{\bf p}')
\nonumber\\
&&-v_{\rm F}({\bf p}-{\bf p}')\}, \nonumber\\
\epsilon_\downarrow(p_y)&=&\epsilon_\uparrow(p_y)
\end{eqnarray}
The energy per particle $E_c(r)$ is given by
\begin{equation}
E_c(r)=\int_{-\pi/4}^{\pi/4}{dp_y\over2\pi}(
\epsilon_\uparrow(p_y)+\epsilon_\downarrow(p_y)).
\end{equation}

\begin{figure}[t]
\begin{center}\leavevmode
\includegraphics[width=0.8\linewidth]{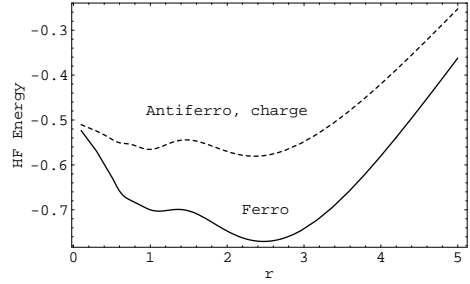}
\caption{Hartree-Fock energy of the ferromagnetic, antiferromagnetic, 
and charge striped state against the period of the stripe $r$. 
}\label{figurename}\end{center}\end{figure}

Let us compare the Hartree-Fock energy obtained in the previous 
calculation for three striped states. 
The energy at $\nu=4+1/2$ is plotted in Fig.~2 as a function of 
the period of the stripe $r$. 
Note that the antiferromegnatic and charge striped states 
have a same energy. 
As seen in the Figure, the ferromagnetic striped state has the 
lower energy than the other states for all period. 
We call this stable ferromagnetic striped state the striped Hall 
ferromagnet. 

We consider the low-lying excitations due to the spontaneous symmetry 
breaking in the striped Hall ferromagnet. 
We approximate the low-lying excitation states for the charge and 
spin excitation as follows, 
\begin{equation}
\vert {\bf k},{\rm charge}\rangle=
\rho({\bf k})\vert0\rangle,\ 
\vert {\bf k},{\rm spin}\rangle=
s_-({\bf k})\vert0\rangle,
\end{equation}
where the state $\vert0\rangle$ is the striped Hall ferromagnet 
state obtained in the Hartree-Fock approximation. 
Note that $s_+({\bf k})\vert0\rangle=0$. 
These excitations correspond to the NG modes of the spontaneous 
breaking of the translation and spin rotation symmetry. 
We call these excitations phonon and spin wave, respectively. 

The excitation energies for these collective mode are given by
\begin{eqnarray}
\Delta_{\rm phonon}({\bf k})&=&{\langle{\bf k},{\rm charge}
\vert(H_{\rm int}-E_0)\vert{\bf k},{\rm charge}\rangle\over 
\langle{\bf k},{\rm charge}\vert{\bf k},{\rm charge}\rangle}
\label{del}\\
\Delta_{\rm spin}({\bf k})&=&{\langle{\bf k},{\rm spin}
\vert(H_{\rm int}-E_0)\vert{\bf k},{\rm spin}\rangle\over 
\langle{\bf k},{\rm spin}\vert{\bf k},{\rm spin}\rangle},
\nonumber
\end{eqnarray}
where $E_0$ is the ground state energy. 

\begin{figure}[t]
\begin{center}\leavevmode
\includegraphics[width=0.8\linewidth]{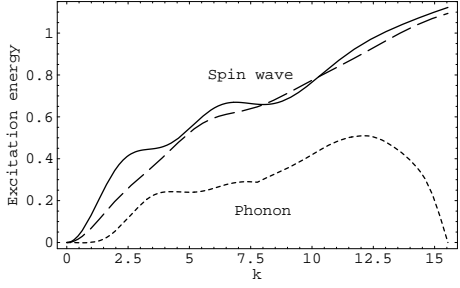}
\caption{The excitation energies for phonon and 
spin wave for $l=2$. 
Solid line shows $\Delta_{\rm spin}(0,k_y)$ and 
dashed line shows $\Delta_{\rm spin}(k_x,0)$. 
Dotted line shows $\Delta_{\rm phonon}(0,k_y)$. 
}\label{figurename}\end{center}\end{figure}

Under the assumption that the ground state is given by the Hartree-Fock 
state for the striped Hall ferromagnet, Eq.~(\ref{del}) 
is written by using the double commutation as
\begin{eqnarray}
\Delta_{\rm phonon}({\bf k})&=&{\langle 0
\vert\left[\rho(-{\bf k}),\left[H_{\rm int},\rho({\bf k})
\right]\right]\vert0\rangle\over 
2\langle0\vert\rho(-{\bf k})\rho({\bf k})\vert0\rangle},\\
\Delta_{\rm spin}({\bf k})&=&{\langle 0
\vert\left[s_+(-{\bf k}),\left[H_{\rm int},s_-({\bf k})
\right]\right]\vert0\rangle\over 
\langle0\vert s_+(-{\bf k})s_-({\bf k})\vert0\rangle},
\nonumber
\end{eqnarray}
and calculated using the following commutation relations 
\begin{eqnarray} 
\left[\rho_*({\bf k}),\rho_*({\bf k}')\right] 
&=&-2i \sin\left({{\bf k}\times{\bf k}'\over4\pi}\right)
\rho_*({\bf k}+{\bf k}'),\nonumber\\ 
\left[s_{*\pm}({\bf k}),\rho_*({\bf k}')\right] &=&-2i 
\sin\left({{\bf k}\times{\bf k}'\over4\pi}\right)
s_{*\pm}({\bf k}+{\bf k}'),\nonumber\\ 
\left[s_{*+}({\bf k}),s_{*-}({\bf k}')\right] &=&
-i\sin\left({{\bf k}\times{\bf k}'\over4\pi}\right)
\rho_*({\bf k}+{\bf k}')\\ 
&&+2\cos\left({{\bf k}\times{\bf k}'\over4\pi}\right)
s_{*z}({\bf k}+{\bf k}')
\nonumber
\end{eqnarray}
and $\left[s_{*\pm}({\bf k}),s_{*\pm}({\bf k}')\right]=0$, where 
$\rho_*({\bf k})=\rho({\bf k})/f_l(k)$ and $s_{*i}({\bf k})=
s_i({\bf k})/f_l(k)$. 

The results for $\nu=4+1/2$ are shown in Fig.~3. 
As seen in this figure, the spectrum for the spin wave has a weaker 
anisotropy than the phonon spectrum. 
We can see $\Delta_{\rm phonon}(k_x,0)=0$ and 
the spectrum for the phonon is highly anisotropic\cite{SMA}. 
On the other hand, it can be seen that 
$\Delta_{\rm spin}(k_x,0)\propto k_x^2$, 
$\Delta_{\rm spin}(0,k_y)\propto k_y^2$, for small $k$. 

In summary we have calculated the Hartree-Fock energy of spin and 
charge striped states at half-filled high Landau 
level in the zero Zeeman energy limit. 
We have shown that 
a ferromagnetic striped Hall state has a lower energy 
than the antiferromagnetic striped state or charge striped state. 
Furthermore, excitation spectra for low-lying excitations, phonon 
and spin wave, have been obtained in the single mode approximation. 

This work was partially supported by the special Grant-in-Aid for 
Promotion of Education and Science in Hokkaido University, and by 
the Grant-in-Aid for 
Scientific Research on Priority area of Research (B) 
(Dynamics of Superstrings and Field Theories, Grant No. 13135201) from 
the Ministry of Education, Culture, Sports, Science and 
Technology, Japan.


\begin{thebibliography}{9}
\bibitem{Lilly}
M. P. Lilly, K. B. Cooper, J. P. Eisenstein, L. N. Pfeiffer,
and K. W. West, Phys. Rev. Lett. {\bf 82}, 394 (1999).
\bibitem{Du}
R. R. Du, D. C. Tsui, H. L. Stormer, L. N. Pfeiffer, K. W. Baldwin,
and K. W. West, Solid State Commun. {\bf 109}, 389 (1999).

\bibitem{Kou}
A. A. Koulakov, M. M. Fogler, and B. I. Shklovskii, Phys. Rev. Lett.
{\bf 76} (1996) 499; M. M. Fogler, A. A. Koulakov, B. I. Shklovskii, 
Phys. Rev. B {\bf 54}, 1853 (1996).
\bibitem{Moe}
R. Moessner and J. T. Chalker, Phys. Rev. B {\bf 54}, 5006 (1996).

\bibitem{Pan}
W. Pan, H. L. Stormer, D. C. Tsui, L. N. Pfeiffer, and K. W. West, 
Phys. Rev. B {\bf 64}, 121305(R) (2001). 
\bibitem{Dem}
E. Demler, D.-W. Wang, S. Das Sarma, and B. I. Halperin, Solid State 
Comm. {\bf 123}, 243 (2002). 
\bibitem{Rez}
E. H. Rezayi, T. Jungwirth, A. H. MacDonald, and F. D. M. Haldane, 
cond-mat/0302271. 

\bibitem{Jun}
T. Jungwirth and A. H. MacDonald, Phys. Rev. B {\bf 63}, 035305 (2000). 
\bibitem{Wan}
D.-W. Wang, S. Das Sarma, E. Demler, and B. I. Halperin, Phys. Rev. 
B {\bf 66}, 195334 (2002). 
\bibitem{Cote}
R. Cote, H. A. Fertig, J. Bourassa, and D. Bouchiha, Phys. Rev. B 
{\bf 66}, 205315 (2002). 

\bibitem{Von}
K. Ishikawa, N. Maeda, T. Ochiai, and H. Suzuki, Physica E 
{\bf 4}, 37 (1999). 
\bibitem{Imo}
K. Ishikawa, N. Maeda, and T. Ochiai, Phys. Rev. Lett. 
{\bf 82}, 4292 (1999); 
N. Maeda, Phys. Rev. B {\bf 61}, 4766 (2000).
\bibitem{SMA}
T. Aoyama, K. Ishikawa, Y. Ishizuka, and N. Maeda, Phys. Rev. B 
{\bf 66}, 155319 (2002). 

\end{thebibliography}
\end{document}